# Multi-Hop Extensions of Energy-Efficient Wireless Sensor Network Time Synchronization

## *Yangyu Wang and Kyeong Soo Kim*

*Department of Electrical and Electronic Engineering, Xi'an Jiaotong-Liverpool University*
*Suzhou, 215123, P. R. China*
*Email: Yangyu.Wang11@student.xjtlu.edu.cn, Kyeongsoo.Kim@xjtlu.edu.cn*

**Keywords:** *Time synchronization, wireless sensor network, multi-hop extension, two-way message exchange*

## Abstract

We present the multi-hop extensions of the recently proposed energy-efficient time synchronization scheme for wireless sensor networks, which is based on the asynchronous source clock frequency recovery and reversed two-way message exchanges. We consider two hierarchical extensions based on packet relaying and time-translating gateways, respectively, and analyze their performance with respect to the number of layers and the delay variations through simulations. The simulation results demonstrate that the time synchronization performance of the packet relaying, which has lower complexity, is close to that of time-translating gateways.

## 1 Introduction

Time synchronization is a critical function for a wireless sensor network (WSN), through which a unified time framework is provided to the whole network for its proper operations including fusing data from different sensor nodes, sharing time-based channel, and coordinated sleep wake-up node scheduling mechanisms [3].

In a typical WSN, there are mainly two types of nodes, which are a head/master node and a sensor/slave node. The master node is equipped with a powerful processor for handling sensory data as a center for data fusion and is supplied with power from outlet. In this case, the master node is connected to existing networks through wireline interfaces. The sensor nodes, however, are battery-powered devices with limited processing capability. It is this asymmetric WSN that we focus on for time synchronization. This asymmetric property indicates that the time synchronization algorithm should be energy-efficient and has low-complexity for message communication at sensor nodes.

There have been proposed several practical time synchronization schemes, such as Reference-Broadcast Synchronization (RBS) [3], Timing-sync Protocol for Sensor Networks (TPSN) [4] and Flooding Time Synchronization Protocol (FTSP) [5]; it is observed that most of the existing time synchronization schemes for WSN are improved or modified versions of RBS, TPSN, or FTSP. These time synchronization schemes operate upon either clustered (e.g., RBS) or hierarchical topology (e.g., TPSN). To properly estimate a time synchronization scheme, both synchronization accuracy and energy efficiency should be taken into account. For our circumstance, we focus on minimizing the energy consumption on battery-powered sensor nodes; the synchronization should be processed mainly in the head node and the algorithm should be simple for sensor nodes.

In [1], an energy-efficient time synchronization scheme has been proposed for asymmetric WSNs, which is based on the asynchronous source clock frequency recovery (SCFR) [2] and reversed two-way message exchange scheme [4]. In this paper, we study the multi-hop extensions of this synchronization scheme and evaluate their performance based on simulation.

## 2 Review of Energy-Efficient WSN Time Synchronization Based on Asynchronous SCFR and Reverse Two-Way Message Exchanges [1]

Based on the asymmetric WSN, the main idea of the proposed time synchronization is to allow sensor nodes to run without synchronizing their clock offsets to that of the reference clock at a head node, while maintaining their clock frequencies tuned to the reference clock. In this case, the clock offset of a sensor node is only estimated at the head node by receiving the message including timestamps from the sensor node based on reverse two-way message exchanges. Therefore, only the clock frequency of sensor node is synchronized to the reference clock but not the clock offset. The frequency ratio between the reference clock and the slave clock can be estimated using SCFR at a sensor node as illustrated in [2], which only requires receiving time-stamped messages from the head node.

The proposed scheme is shown in Figure 1 along with the conventional two-way message exchange scheme. As shown in the figure, for the conventional scheme, the slave (i.e., the sensor node) firstly sends request to the master (i.e., the head node), and then the master responses message immediately to the slave after receiving the request. After receiving the response, the slave takes time to bundle measurement data and reports to the master. With regard to the proposed reverse two-way message exchange scheme, the master node periodically send request to the slave at first; once the slave has measurement data to report, it replies the response with bundled measurement data and corresponding timestamps to the master.

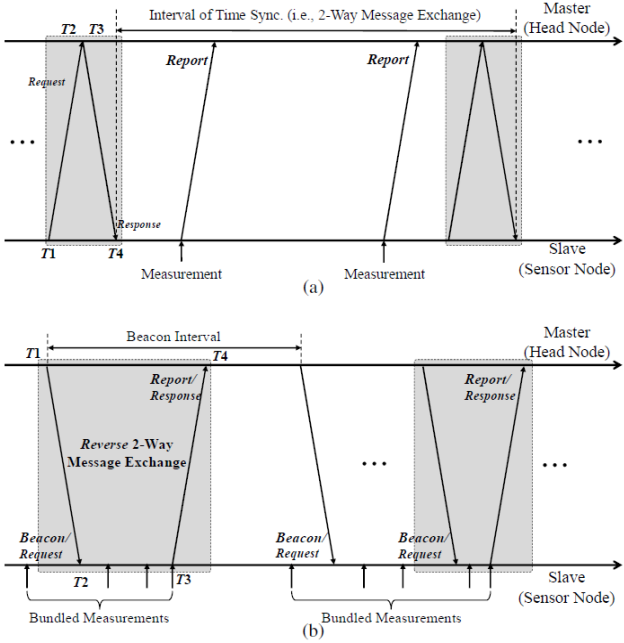

Figure 1. Comparison of two-way message exchange scheme (a) Conventional two-way message exchange scheme introduced in [4]; (b) Proposed reverse two-way message exchange scheme introduced in [1]

With regard to a clock model, among the asymmetric WSN, there is one head node and several sensor nodes, and all of them are equipped with independent hardware clocks based on quartz crystal oscillators. In this case, time $t$ at the head node is treated as a global reference clock. Thus, the hardware time at a slave, says $T$, can be modeled according to [6] as following:

$$T = R \cdot t + \theta \qquad (1)$$

where $R$ and $\theta$ are clock frequency ratio (also called clock skew in some literatures, which is the ratio of the head clock frequency to that of the sensor) and clock offset, respectively.

Since the time synchronization among sensor nodes are independent with one another, one head node to one sensor node time synchronization can be extended to one head node to a number of slaves independently. Therefore, we consider one master and one slave for two-way massage exchange in this model.

In addition, a logical clock is applied in the synchronization operation at a sensor node, which is a function of a physical clock. For evaluating a logical clock, the frequency or offset adjustment by SCFR is taken into account in this case. Therefore, we define $\mathcal{T}$ as the logical clock at the sensor node, and it can be modelled as a piecewise linear function as follows: For $t_k < t \leq t_{k+1}$ $(k = 0,1,\dots)$,

$$\mathcal{T}\big(T(t)\big) - \mathcal{T}\big(T(t_k)\big) = \frac{T(t) - T(t_k)}{\widehat{R}_k} - \widehat{\theta}_k \qquad (2)$$

where $t_k$ is the reference time when $k$th synchronization occurs, and $\widehat{R}_k$ and $\widehat{\theta}_k$ are estimated clock frequency ratio and clock offset from the $k$th synchronization. In this case, $\widehat{\theta}_k$ is set to 0 if the synchronization is only for frequency, and $\widehat{R}_k$ is set to 1 if the synchronization is only for offset. Particularly, since only the clock frequency of a sensor node is synchronized to the reference clock but not the clock offset in the proposed scheme, $\widehat{\theta}_k$ is set to 0 in the logical clock model (i.e., equation (2)).

As described in [4], the clock offset can be evaluated based on estimated timestamps (i.e., T1, T2, T3, T4 in Figure 1(b)) as follows:

$$\widehat{\theta} = \frac{(T2 - T1) - (T4 - T3)}{2} \qquad (3)$$

## 3 Multi-hop Extension of the Proposed Scheme

The proposed time synchronization is illustrated as a single-hop case in [1], which is one Head node communicates with several Sensor nodes. However, this scheme can be extended to multi-hop cases, which is based on a hierarchical structure.

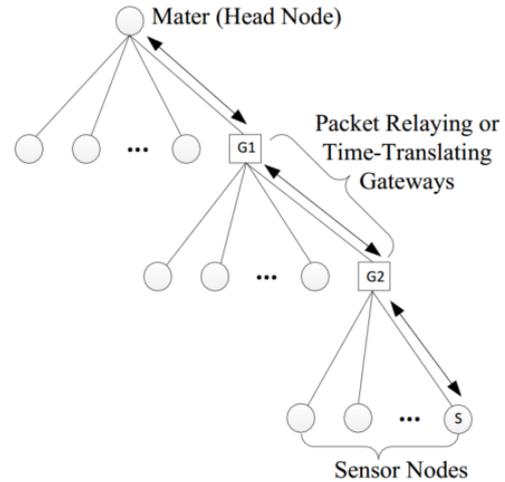

Figure 2. Multi-hop case of proposed time synchronization structure [1]

The Figure 2 shows the multi-hop extension of the proposed time synchronization scheme. Based on a hierarchical structure, the node in a lower layer than the Head node (except for Sensor nodes) can act as both a master node for lower layer, and a normal slave node for its higher layer. In this case, these nodes can be treated as *Gateway* nodes. For example, G1 is the slave node for the Head node, while it is the master node for G2 and other nodes belonging to its next lower layer.

Based on this structure, there are two time synchronization approaches for multi-hop extension, which are:

- Packet relaying
- Time-translating gateways

In the following context, both approaches will be illustrated respectively. We use the *estimation of measurement time* as a major performance measure for time synchronization in the performance evaluation through simulations in the next section.

In the following explanation of multi-hop extension structure, we illustrate the relation between two nodes synchronizing with each other at the same layer as *master* and *slave*, where master node sends the request beacons to slave nodes continuously. We define the node initiating the whole synchronization as the *Head* node (e.g., Head node in Figure 2) and the nodes measuring the data as the *Sensor* node (e.g., Sensor nodes in Figure 2). The nodes between Head node and Sensor node, which are responsible for relaying, are *Gateway* nodes. With regard to the order of multi-hop layers, we set Head node and its slave nodes as the highest layer (e.g., Master and G1 in Figure 2), which is numbered as *Layer 1*, and we set Sensor nodes and its master node as the lowest layer (e.g., Sensor node and G2 in Figure 2), which is numbered as *Layer N,* N $\geq$ 1.

## 3.1 Packet Relaying

For this packet relaying synchronization structure, the gateway node represents the relay for the message exchange between Sensor node and Head node. In this case, the message (e.g., timestamps) from Sensor node is transmitted through Gateway nodes to Head node without any change.

For example of measurement data transmission, in Figure 2, G2 simply passes the received data (i.e., response beacon) from S to G1. Afterwards, G1 receives this message and then sends it to Head node without any change. At last, the Head node will evaluate the clock offset between its clock and corresponding Sensor node clock, and thus applies the time synchronization for the WSN.

In addition, using the example in Figure 2, the time diagram of multi-hop extension using packet relaying approach in real case is shown in Figure 3.

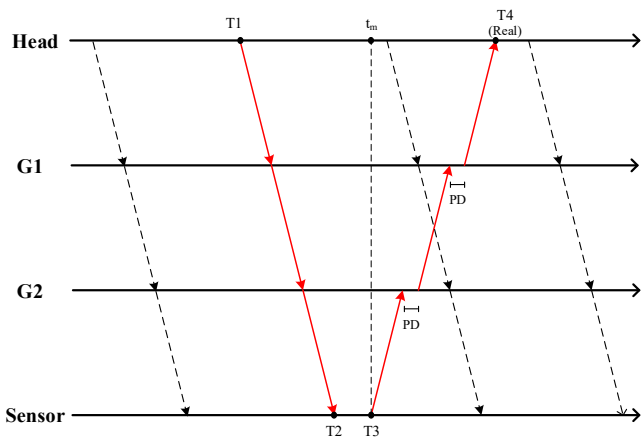

Figure 3. Time diagram of multi-hop (3 hierarchies) extension using Packet Relaying

In this case, the Head node initiates the message exchange and sends the request beacon to it slave node continuously. Then, the Gateway node sends the request beacon to its slave nodes as soon as it achieves the request from its master. Meanwhile, the Sensor node achieves the measurement data and sends it with the time stamps to its master node. In this case, the Gateway node transfers the message from its slave node to master node once it receives the response beacon from its slave node. Finally, the measurement data with timestamps will be transferred from Sensor node to Head node without any change.

The time stamps in Figure 3 are explained in the following.

- *T1*: reference clock time that the request beacon departures from Head node, which is generated based on Poisson process;

- *T2*: logical clock time after synchronization based on request arrival time at Sensor node;

- *T3*: updated measurement time (in logical clock) based on logical model and estimated frequency ratio between Head node and Sensor node, which is also the time response departures from Sensor node;

- *T4*: reference clock time that measurement message (i.e., response beacon) arrives at Head node.

Particularly, it should be noticed that, at each layer, the Gateway node sends request to it slave immediately after receiving the request from it higher layer, which means the request arrival time (from higher layer) and request departure time (to lower layer) are same at each Gateway node. It actually shows that the request beacon sending times are correlated in the whole network.

Especially, there is a processing delay (marked as PD in Figure 3) between the response arrival time from slave and departure time to master at Gateway node. Moreover, in reality, the processing delay (i.e., queueing and MAC operation according to [1]) will significantly affect the accuracy of synchronization. Therefore, it suggests that the processing delay should be compensated at Head node.

Therefore, the equivalent compensated time diagram can be shown as following Figure 4.

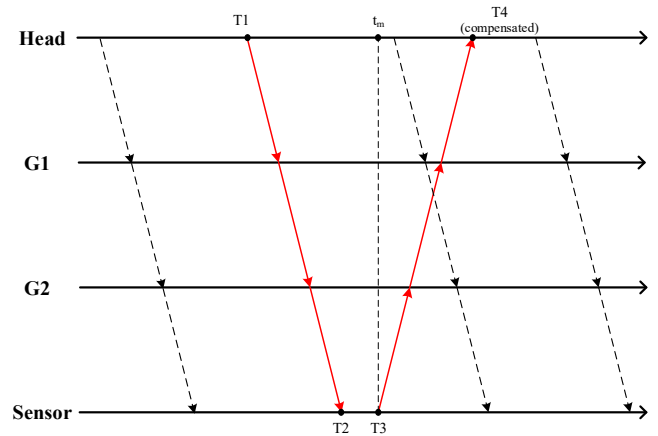

Figure 4. Equivalent compensated time diagram of multi-hop (3 hierarchies) extension using Packet Relaying

In this case, for simulation of measurement time estimation, we can actually consider the above four timestamps (i.e., T1, T2, T3 and T4 (compensated)) and only evaluate the clock offset between Head node and Sensor node and thus achieve the estimated measurement time. Therefore, the Gateway node has no effect on the measurement time estimation.

## 3.2 Time-translating Gateways

For this time-translating gateways synchronization structure, the time synchronization for each hierarchy is independent. In this regard, the time synchronization will be operated at each layer (i.e., between gateway nodes, gateway node and sensor node, or head node and gateway node), which starts from lowest layer (i.e., Sensor node and Gateway node).

For the simulation of measurement time estimation, at lowest layer (i.e., sensor node and gateway node), the measurement time is estimated as single-hop case illustrated before. Furthermore, the higher layer will also estimate the measurement time based on the estimated measurement time in lower layer and the current layer synchronization conditions (i.e., the timestamps and logical clocks in its own layer). In this case, the previous estimated measurement time can be treated as the hardware clock time in lower layer applied in logical clock model (i.e., equation (2)). Therefore, after several times of synchronization, the final version of estimated measurement time will be achieved at Head node

For example, in Figure 2, between G2 and Sensor, the proposed synchronization will be operated as single-hop case, where G2 achieves the estimated measurement time, says $t_{m\_v1}$. Afterwards, based on synchronization conditions between G1 and G2, as well as $t_{m\_v1}$, the estimated measurement time, says $t_{m\_v2}$, can be achieved at G1. At last, similarly, based on synchronization conditions between Head and G1, as well as $t_{m\_v2}$, the estimated measurement time, says $t_{m\_v3}$, can be achieved at Head node. In this case, $t_{m\_v3}$ is the final estimated measurement time for this synchronization.

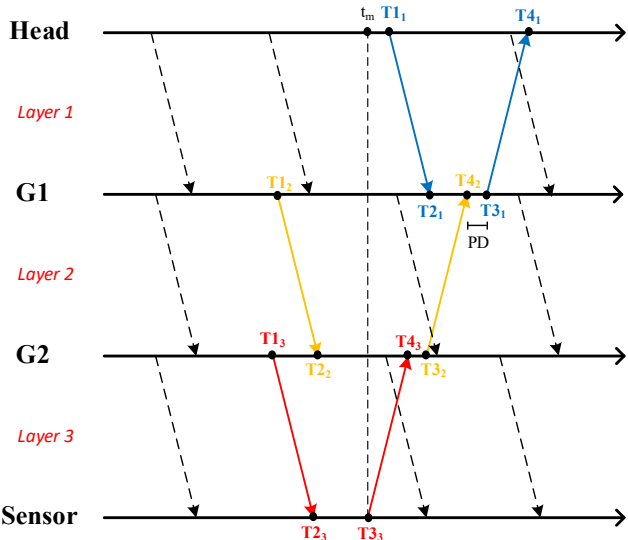

Figure 5. Time diagram of multi-hop (3 hierarchies) extension using Time-translating Gateways

In addition, using the example in Figure 2, the time diagram of multi-hop extension using time-translating gateways approach is shown in Figure 5. In this case, the time synchronizations conducted in different layers are marked in different colors as well as the corresponding timestamps.

With this regard, we mean layer i synchronization by measurement time estimation of node at layer i. In Figure 5, we set $Tn_i$ by the number n timestamps for layer i synchronization, for example, $T4_2$ is the fourth timestamp (i.e., arrival time of measurement message) for layer 2 synchronization between G1 and G2.

For estimating measurement time at layer i synchronization, the formula can be constructed as following.

$$t_{m\_est}^{i} = \frac{(t_{m\_est}^{i+1} - T_{2*}^{i})}{\hat{R}^i} + T_2^i - \hat{\theta}^i \qquad (4)$$

- $t_{m\_est}^{i+1}$ : the estimated measurement time from the operation of layer i+1 (lower layer) synchronization;
- $\hat{R}^i$ : the estimated frequency ratio of layer i synchronization (i.e., the frequency ratio between the master node and slave node at layer i) based on SCFR;
- $T_2^i$ : the logical clock time after skew-correction based arrival time of request beacon for layer i synchronization;
- $T_{2*}^i$: the arrival time of request beacon in hardware clock for layer i synchronization with respect to slave node;
- $\hat{\theta}^i$: The estimated clock offset of layer i synchronization.

Particularly, since the measurement time is estimated at each layer (except for sensor node), there should be processing delay during the translation interval. For example, there is an interval for processing delay between $T4_2$ and $T3_1$ in Figure 5, which is marked as PD. Unlike the packet relaying approach, the PD of time-translating gateways will not affect the final estimation result. It is due to the independency of evaluation of clock offset for each layer.

Moreover, it should be noticed that, not like the packet relaying extension, the request beacon sending time at each gateway nodes and master are not related with each other in the whole network. They are independent for different layers. In this case, at each layer, the master can send the request to slave for any time, which is not like packet relaying that the request beacons should be sent from Head node to Sensor node one layer by one layer in sequence.

## 4 Simulation Results

We carry out the simulation to investigate the performance of multi-hop extension with regard to the number of hops/layers. Two approaches (i.e., packet relaying and time-translating gateways) are investigated and their performance is compared through simulation. We quantified the performance of the time synchronization with the accuracy of estimation of measurement time; the mean square error (MSE) between measurement time in reference time and the estimated one is treated as the measure for comparison in simulation. Therefore, the MSE of estimated measurement time with regard to the number of layers is investigated.

Below are basic assumptions for all simulation experiments:

- The observation period (simulation period) was set to be 3600 seconds;
- The number of measurement during the observation period is 100;
- The distance between nodes within same layers is 100 meters;
- The propagation speed of beacon is $3 \times 10^8$ m/sec;
- The measurement time in reference clock is generated randomly based on Poisson process;
- The number of bundled measured data is only set to 1, which means Sensor node sends message to its master once it achieves the measurement data.

The parameters in clock model (i.e., equation (1)) are set as variables for comparison in the simulation, and the default settings are as follows:

- The number of layers for investigation is from 1 to 20, which means the number of gateway nodes is from 0 to 19;
- The frequency ratio (i.e., R in equation (1)) between master and slave in each layer is randomly generated in range of 1 - 100 ppm and 1 + 100 ppm, which follows the normal distribution;
- The clock offset (i.e., θ in equation (1)) between master and slave at each layer is randomly generated in range of -1 and +1 second, which follows the normal distribution;

The standard deviation of random delay during the propagation is set as $10^{-9}$ second (i.e., 1 ns) as default; this random delay mainly indicates the random processing delay for transmitting and receiving messages at nodes. It is generated randomly during each transmission, which follows the normal distribution. It is treated as a variable for comparison during the simulation.

We assume the request/response beacons can be transmitted and received successfully, where no operation problem (e.g., retransmission or big processing delay) is included. In this case, we ignore the influence of MAC operations and queueing problems at Gateway nodes during time synchronization.

We focus on the procedure of time synchronization. Thus we can achieve the layered relation between the nodes within WSN (i.e., the layered structure as Figure 3 and Figure 5 show) during the simulation, and the problems related to message transmission within the network (e.g., choice of route of transferring message or choice of Gateway nodes) are not concerned in the simulation.

The simulation results for the performance of multi-hop extension using two approaches are shown in Figure 6. It can be found that the MSE of estimated measurement time increases with the increase in number of layers in WSN. It means the performance of time synchronization using packet relaying becomes worse with more number of hops. The MSE of estimated measurement time rises rapidly until the number of layers becomes 5; the MSE increases by almost 5 times as the number of layers grows from 1 to 5. Afterwards, for the number of layers going up to 20, the growth rate slows down, and the curve shows a linear behaviour. Then, the MSE reaches $10^{-17}$ when the number of layers is 20, where it increases by nearly 4 times with the number of layers growing from 5 to 20. In this case, we infer that the MSE will increase more slowly as we add more layers and possibly saturates.

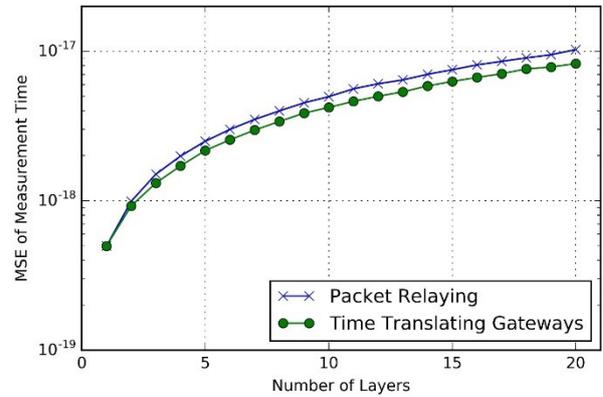

Figure 6. The comparison between two multi-hop extension approaches for default setting

The Figure 6 also indicates that, for single hop (i.e., the number of layers is 1), two extension approaches show almost same performance. However, for further extension (number of layers > 1), the MSE of estimated measurement time using time-translating gateways is lower than the one using packet relaying, which means the performance of time-translating gateways is slightly better than packet relaying. It can be explained that, the time synchronization is operated at each layer for the approach of time-translating gateways, but the approach of packet relaying only estimates measurement time once based on the time stamps in highest layer (i.e., Head node) and lowest layer (i.e., Sensor node). In this case, for time-translating gateways, the error of estimation can be corrected at each layer, rather than packet relaying corrects the error once after the errors are accumulated during transmission across all layers.

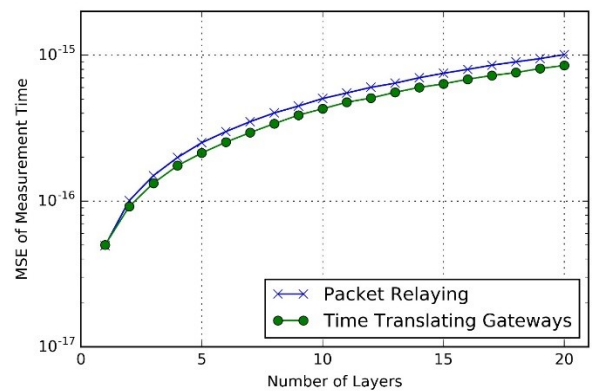

Figure 7. The comparison between two multi-hop extension approaches for increasing Standard Deviation of Random Delay

To investigate the impact of noise variation on time synchronization, we increase the standard deviation of random delay from $10^{-9}$ to $10^{-8}$, and then compare the performance between the two approaches. The results are shown in Figure 7. It can be found that, for higher standard deviation of random delay, the performance of both approaches are much worse than that with lower standard deviation.

We also carried out simulation with different values of the standard deviation of clock offset and frequency ratio.[1] The simulation results show that the variance of distribution of clock offset and frequency ratio does not affect the performance of multi-hop extension of time synchronization for both two approaches.

## 5  Conclusions

In this paper, we focus our study on the asymmetric WSN which contains mainly two kinds of nodes, i.e., a master/head node and a slave/sensor node. The master node is equipped with a powerful processor for handling sensory data and supplied power from outlet. The sensor nodes, on the other hand, are battery-powered devices with low processing capability. This requires that the time synchronization algorithm should be energy-efficient and has low-complexity for message communication at sensor nodes. We have discussed the energy-efficient time synchronization scheme for asymmetric WSNs based on structure, clock model, frequency ratio estimator, and simulation.

We have presented the theoretical design of multi-hop extension for the time synchronization scheme proposed in [1]. We have shown two approaches for its hierarchical extension, which are packet relaying and time-translating gateways. For packet relaying, the time synchronization is operated base on time stamps at Head node and Sensor node, where Gateway node only relays the request/response message without any change in the middle. As for time-translating gateways, the operation of time synchronization is applied at each layer between master and slave, and final estimated result is achieved at Head node. In this case, the complexity of time-translating gateways is much higher than packet relaying.

We have developed detailed simulation models and carried out a comparative analysis of the two extension approaches through simulation. The simulation results demonstrate that the time-translating gateways perform better than packet relaying for the cases considered. We also find that the variances of distribution of clock offset and frequency ratio do not much affect the performance of both multi-hop extension approaches, while the variance of delay distribution significantly affects the accuracy of time synchronization.

For future work, we plan to carry out mathematical analysis of the simulation results, where we will investigate the reason of the trend of time synchronization performance, which results in the strategy of improving the multi-hop extension approaches.

Another area of research for further work is more detailed simulation with realistic assumptions. For example, MAC operations, queueing and scheduling delays may be introduced in the simulation. In this case, the advanced simulation software with a full network protocol stack (e.g., OMNeT++/INET) can be applied.

## References

[1] K. S. Kim, S. Lee and E. G. Lim, "On energy-efficient time synchronization based on source clock frequency recovery in wireless sensor networks," in *Proceeding of International Conference on Information, System and Convergence Applications (ICISCA) 2015*, Kuala Lumpur, Malaysia, 2015.

[2] K. S. Kim, "Asynchronous Source Clock Frequency Recovery through Aperiodic Packet Streams," *IEEE Communications Letters,* vol. 17, no. 7, pp. 1455-1458, 2013.

[3] J. Elson, L. Girod and D. Estrin, "Fine-grained network time synchronization using reference broadcasts," *Acm Sigops Operating Systems Review,* pp. 147-163, 2002.

[4] S. Ganeriwal, R. Kumar and M. B. Srivastava, "Timing-sync Protocol for Sensor Networks," in *International Conference on Embedded Networked Sensor Systems*, 2004.

[5] M. Maróti, B. Kusy, G. Simon and Á. Lédeczi, "The Flooding Time Synchronization Protocol," in *International Conference on Embedded Networked Sensor Systems*, Baltimore, 2004.

[6] R. T. Rajan and A.-J. v. d. Veen, "Joint ranging and clock synchronization for a wireless network," in *Computational Advances in Multi-Sensor Adaptive Processing (CAMSAP), 2011 4th IEEE International Workshop on*, 2011.

[7] Y. C. Wu, Q. Chaudhari and E. Serpedin, "Clock synchronization of wireless sensor networks," *IEEE Signal Processing Magazine,* vol. 28(1), pp. 124-138, 2011.

---

[1] Due to the limited space, the results are not shown here.